\documentclass[onecolumn,amsmath,showpacs,superscriptaddress,12pt]{revtex4-2}
\usepackage{graphicx}
\usepackage{dcolumn}   
\usepackage{bm}        
\usepackage{amssymb}   
\usepackage{amsmath}
\usepackage[utf8]{inputenc}
\usepackage{color}
\usepackage{array}
\usepackage{physics}
\usepackage{makecell,booktabs,multirow}
\usepackage{stackengine}
\usepackage{hyperref}
\hypersetup{
    colorlinks=true,
    linkcolor=blue,
    citecolor=blue,
    filecolor=magenta,      
    urlcolor=cyan,
}

\begin{document}

\title{Friedel oscillations in a two-dimensional electron gas and monolayer graphene with a non-Coulomb impurity potential}

\author{Levente Máthé}
\email[Corresponding author: ]{levente.mathe@itim-cj.ro}
\affiliation{National Institute for Research and Development of Isotopic and Molecular Technologies, 67-103 Donat, 400293 Cluj-Napoca, Romania}

\author{Ioan Grosu}
\affiliation{Faculty of Physics, Babeș-Bolyai University, 1 Kogălniceanu, 400084 Cluj-Napoca, Romania}

\begin{abstract}
We study Friedel oscillations in a two-dimensional non-interacting electron gas and in a monolayer graphene in the presence of a single impurity. The potential generated by the impurity is modeled using a non-Coulomb interaction ($\sim r^{-\eta}$). The charge carrier density deviation as a function of distance from the impurity is calculated within the linear response theory. Our results show that, in both a two-dimensional non-interacting electron gas and graphene, the phase of charge carrier density oscillations remains unaffected by the parameter $\eta$, which characterizes the non-Coulomb nature of the interaction, at large distances from the impurity. The parameter $\eta$ influences only the amplitude of the oscillations in this regime. The results for an impurity modeled by Coulomb-like potential ($\eta = 1$) are recovered in both cases. 
\end{abstract}


\maketitle

\section{Introduction}
\label{sec:I}
Friedel oscillations are a fundamental phenomenon in condensed matter physics that describe the oscillatory behavior of the electron density in the vicinity of a perturbing impurity or defect in a material. Friedel oscillations originate from the quantum behavior of electrons and their scattering from localized perturbations, such as charged impurities, in the crystal lattice~\cite{Friedel1952,Villain2015}. When an impurity is introduced into a conductive material, it creates a potential that scatters the conduction electrons. The interference of the scattered electron waves results in oscillations in the electron density around the impurity. The oscillations decay with the distance from the impurity and the frequency is related to the Fermi wave vector. The amplitude of the oscillations depends on the strength of the impurity potential. In a typical three-dimensional metal, the Friedel oscillations decays as $r^{-3}$, where $r$ is the distance from the impurity, while in two-dimensions the decay of Friedel oscillations is slower, as $r^{-2}$~\cite{Giuliani2005a}. The reduced dimensionality affects the scattering potential and the nature of the electron interactions. The situation differs for graphene~\cite{Wunsch2006,Cheianov2006,Pyatkovskiy2008,Bacsi2010,Thakur2017,Katsnelson2020,Elsayed2023,Lin2024}, a single layer of carbon atoms arranged in a honeycomb lattice. Graphene exhibits unique electronic properties due to its Dirac cone band structure. In this case, the oscillations decay much faster than in a conventional two-dimensional electron gas, a faster decay attributed to the suppression of backscattering due to the chiral nature of the Dirac electrons.

Friedel oscillations have been extensively studied in the literature for a one-dimensional non-interacting electron gas by considering the presence of a single impurity~\cite{Giuliani2005}, two-impurities~\cite{Grosu2008,Mathe2023}, many-impurities~\cite{Tugulan2008} or a chain of dense impurities~\cite{Grosu2007}.
The effects of interactions, temperature, transport and spin-related phenomena on Friedel oscillations in one-dimensional systems have been extensively studied in previous works~\cite{Fabrizio1995,Gorczyca2007,Gorczyca2009,Soffing2009,Ziani2012,Gambetta2015,Kylanpaa2016,Ziani2021}.
The problem of Friedel oscillations in superconductors has also been theoretically examined using the Bogoliubov-de Gennes formalism~\cite{Machida2003,Lauke2018,Stosiek2022}.
Recent studies have explored Friedel oscillations in the case of a non-Hermitian imaginary impurity employing non-Hermitian linear response theory~\cite{Dora2021,Sticlet2024}.
The problem of Friedel oscillations has been studied in two- and three-dimensional electron Fermi liquids with both short- and long-range impurity potentials using linear response theory~\cite{Giuliani2003,Simion2005}. 
The Friedel oscillations have been investigated in different systems with correlated lattice electrons in the presence of a single impurity, many impurities or defects using dynamical mean-field theory~\cite{Chatterjee2015,Chatterjee2019,Chatterjee2022}.
Recently, Friedel oscillations have been theoretically studied in two-dimensional topological materials with Mexican-hat band dispersion~\cite{Sablikov2025}.
The Friedel oscillations were experimentally demonstrated in different materials by using scanning tunneling microscopy~\cite{Crommie1993,Hofmann1997,Kanisawa2001,Ono2009,Dutreix2019,Chen2021,Yin2023}, Mössbauer spectroscopy~\cite{Mitsui2020,Mitsui2021}, nuclear magnetic resonance~\cite{Berthier1978,Yamani2006} or X-ray diffraction~\cite{Rouziere2000,Czoschke2005} measurements. 
Many-particle systems with non-Coulomb interactions play a significant role in both theoretical and experimental condensed matter and atomic physics. Such behavior is encountered in systems as ultra-cold atomic and molecular gases (dipolar interacting molecular gases, trapped ions, Rydberg atoms), where long-range interactions dominate, non-Coulomb behavior in heterostructures and at interaction between an interface and an adjacent electrolyte solution~\cite{Katsnelson2006,Gabovich2019,Vangara2019,Zhang2022,Defenu2024,Radzihovsky2024}.

In this work, we investigate Friedel oscillations induced by a single impurity, modeled using a non-Coulomb interaction of the form $r^{-\eta}$, in two systems: (i) two-dimensional non-interacting electron gas and (ii) monolayer graphene. The primary objective of this study is to analyze the impact of the parameter $\eta$, which describes the non-Coulomb character of the interaction, on the physics of Friedel oscillations. 
\section{Theory}
\label{sec:II}
In our model, the perturbing potential of a single impurity in two-dimensional non-interacting electron gas or monolayer graphene is characterized by a non-Coulomb potential, expressed as~\cite{Ene2025}:
\begin{equation}
V(r)=\frac{A}{r^\eta},
\label{eq:1}
\end{equation}
where $\eta$ characterizes the non-Coulomb character of the interaction. The dimensionality of $A$ is such that Eq.~\eqref{eq:1} has the dimension of energy.

In the linear response theory, the dependence of charge carrier density on distance $r$, measured from the perturbing impurity, is given by~\cite{Simion2005,Giuliani2005a}:
\begin{equation}
n(r)=\int \frac{d^2 q}{(2\pi)^2}  e^{i \vec{q}\cdot\vec{r}}\, V(q) \, \chi_0(q),
\label{eq:2}
\end{equation}
where $V(q)$ is the Fourier transform of the perturbing impurity potential $V(r)$, which reads
\begin{equation}
	V(q)=\int d^2 r \, e^{-i \vec{q}\cdot\vec{r}}\, V(r) = \frac{2^{2-\eta} A \, [\Gamma(1-\frac{\eta}{2})]^2 \sin(\frac{\pi \eta}{2})}{q^{2-\eta}},
	\label{eq:3}
\end{equation}
a result valid for $\frac{1}{2}<\eta < 2$, where $\Gamma(z)$ stands for the gamma function. In addition, $\chi_0(q)$ represents the static Lindhard response function which for two-dimensional non-interacting electron gas is ($\hbar=1$)~\cite{Giuliani2005a,Simion2005} 
\begin{equation}
\chi_0^{2d}(q) = -\frac{m}{\pi}\bigg [ 1-\theta(q-2k_F)\sqrt{1-\Big( \frac{2k_F}{q} \Big)^2 }\, \bigg],
\label{eq:4}
\end{equation}
where $m$ and $k_F$ stand for the electron mass and Fermi momentum, and $\theta(x)$ is the Heaviside's step function. 
In the case of a monolayer graphene the static Lindhard response function is given by ($\hbar=1$)~\cite{Hwang2007}
\begin{equation}
	\chi_0^{gr}(q) = D(E_F)\bigg \{ 1-\theta(q-2k_F) \bigg [ \frac{1}{2} \sqrt{1-\Big (\frac{2k_F}{q}\Big )^2} +\frac{q}{4k_F} \arcsin(\frac{2k_F}{q}) -\frac{\pi q}{8k_F} \bigg ] \bigg \},
	\label{eq:5}
\end{equation}
where $D(E_F) = g_s g_v E_F/2\pi\gamma^2$ is the density of states in graphene at the Fermi level $E_F = \gamma \, k_F$ with $\gamma$ being a band parameter, i.e., the graphene Fermi velocity. The Fermi momentum is given by $k_F=\sqrt{4\pi n_c/g_v g_s}$, where $g_v = 2$ and $g_s = 2$ are the valley and spin degeneracies, respectively. Here, $n_c$ represents the charge carrier (electron/hole) density. 

In the following, we determine the charge carrier density $n(r)$ as a function of distance from the impurity for the cases listed above.

\textit{Case (i): Two-dimensional non-interacting electron gas}

Substituting Eq.~\eqref{eq:4} into Eq.~\eqref{eq:2}, one obtains
\begin{equation}
n(r)=-\frac{m}{\pi}\bigg [ \int \frac{d^2 q}{(2\pi)^2}  e^{i \vec{q}\cdot\vec{r}}\, V(q)-\int \frac{d^2 q}{(2\pi)^2}  e^{i \vec{q}\cdot\vec{r}} \, V(q)\,\theta(q-2k_F)\sqrt{1-\Big( \frac{2k_F}{q} \Big)^2 } \,\bigg ]. 
\label{eq:6}
\end{equation}
By expressing the two-dimensional integral in Eq.~\eqref{eq:6} in polar coordinates, together with Eq.~\eqref{eq:3}, we obtain
\begin{equation}
	\begin{split}
n(r)=-\frac{m A}{\pi}&\bigg [ \frac{1}{r^\eta}-\frac{2^{2-\eta}[\Gamma(1-\frac{\eta}{2})]^2 \sin(\frac{\pi \eta}{2})}{(2\pi)^2}\\
&\times\int_{0}^{\infty} \frac{dq\, q}{q^{2-\eta}} \,\int_{0}^{2\pi} d\phi \, e^{i q\,r \cos \phi}\, \theta(q-2k_F) \,\sqrt{1-\Big( \frac{2k_F}{q} \Big)^2} \,  \bigg ],
	\end{split}
\label{eq:7}
\end{equation}
where $\phi$ is the polar angle. Thus, Eq.~\eqref{eq:7} reduces to
\begin{equation}
	n(r)=-\frac{m A}{\pi}\bigg [ \frac{1}{r^\eta}-\frac{2\,[\Gamma(1-\frac{\eta}{2})]^2 \sin(\frac{\pi \eta}{2})}{2^\eta\, \pi}\int_{2k_F}^{\infty} dq\, q^{\eta-1} \, J_0(qr) \,\sqrt{1-\Big( \frac{2k_F}{q} \Big)^2}\, \bigg ], 
	\label{eq:8}
\end{equation}
where $J_\nu(qr)$ represents the Bessel function of the first kind with $\nu=0$. By introducing the variable $x = q/2k_F$, one finds
\begin{equation}
n(r)=-\frac{m A}{\pi}\bigg [ \frac{1}{r^\eta} -\frac{2 \, k_F^\eta \, [\Gamma(1-\frac{\eta}{2})]^2 \sin(\frac{\pi \eta}{2})}{\pi} \int_{1}^{\infty} dx\, x^{\eta-2} \, J_0(2k_F r x) \,\sqrt{x^2-1} \bigg ].
\label{eq:9}
\end{equation}
Denoting by $I_0$ the integral from Eq.~\eqref{eq:9}, we obtain
\begin{equation}
\begin{split}
I_0 &=\int_{1}^{\infty} dx\, x^{\eta-2} \, J_0(2k_F r x) \,\sqrt{x^2-1}= \frac{\sqrt{\pi}}{4(k_F r)^\eta}\\
&\times \Bigg [ \frac{2\,\Gamma(\frac{\eta}{2})\,{}_1F_2(-\frac{1}{2};1-\frac{\eta}{2},1-\frac{\eta}{2};-(k_F r)^2)}{\sqrt{\pi}\,\Gamma(1-\frac{\eta}{2})} + \frac{\Gamma(-\frac{\eta}{2})\,{}_1F_2(\frac{\eta -1 }{2};1,1+\frac{\eta}{2};-(k_F r)^2)}{(k_F r)^{-\eta}\,\Gamma(\frac{3-\eta}{2})} \Bigg ],
\end{split}
\label{eq:10}
\end{equation}
where ${}_1F_2(a;b,c;z)$ is the generalized hypergeometric function and the result is valid for $\eta<\frac{3}{2}$. Applying now the identities~\cite{Gradshteyn2007}:
\begin{equation}
	\begin{aligned}
		\Gamma(1-z) &= \frac{\pi}{\Gamma(z)\sin(\pi z)},\\
		\Gamma(1+z) &= z \,\Gamma(z),\\
	\end{aligned}\label{eq:11}
\end{equation}	
and relation~\eqref{eq:10}, the dependence of electron density on distance $r$, given by Eq.~\eqref{eq:9}, can be written as
\begin{equation}
\begin{split}
n(r)=&-\frac{m k_F^\eta A}{\pi} \frac{1}{(k_F r)^\eta} \Bigg \{ 1 -\bigg [ {}_1F_2\Big (-\frac{1}{2};1-\frac{\eta}{2},1-\frac{\eta}{2}; -(k_F \, r)^2\Big )\\
&-\frac{\pi^{\frac{5}{2}}\,(k_F r)^\eta }{\eta \, \Gamma(\frac{3-\eta}{2})[\Gamma(\frac{\eta}{2})]^3 \sin^2(\frac{\pi \eta}{2})}\, {}_1F_2\Big (\frac{\eta-1}{2}; 1,1+\frac{\eta}{2}; -(k_F \, r)^2\Big ) \bigg ] \Bigg \},
\end{split} 
\label{eq:12}
\end{equation}
with the condition $\frac{1}{2}<\eta<\frac{3}{2}$. We introduce $\delta n(r)$ as the oscillatory term of $n(r)$ in Eq.~\eqref{eq:12}, and obtain
\begin{equation}
\begin{split}
\delta n(r)&=\frac{m k_F^\eta A}{\pi} \frac{1}{(k_F r)^\eta} \bigg [ {}_1F_2\Big (-\frac{1}{2};1-\frac{\eta}{2},1-\frac{\eta}{2}; -(k_F \, r)^2\Big )\\
&-\frac{\pi^{\frac{5}{2}}\,(k_F r)^\eta }{\eta \, \Gamma(\frac{3-\eta}{2})[\Gamma(\frac{\eta}{2})]^3 \sin^2(\frac{\pi \eta}{2})} \, {}_1F_2\Big (\frac{\eta-1}{2}; 1,1+\frac{\eta}{2}; -(k_F r)^2\Big ) \bigg ].
\end{split} 
\label{eq:13}
\end{equation}

\begin{figure}
	\includegraphics[width=14cm,keepaspectratio]{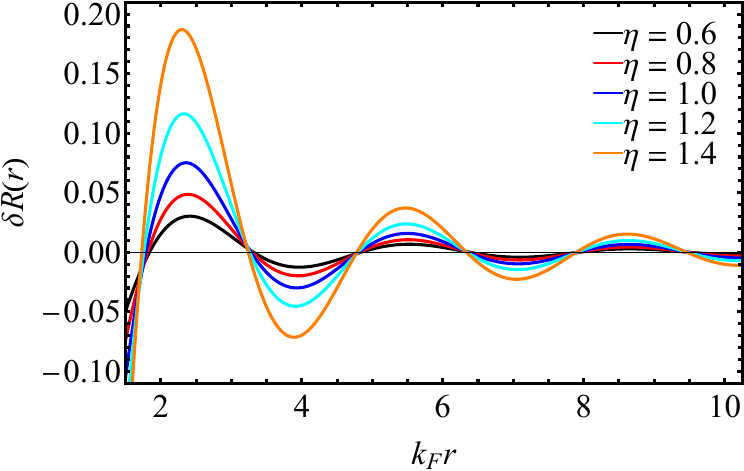}
	\caption{The normalized oscillatory term of the electron density $\delta R(r) = \pi\delta n(r)/(m k_F^\eta A)$ as a function of the dimensionless distance $k_F r$ for two-dimensional non-interacting electron gas with different values of the number $\eta$, and $k_F=1$.}
	\label{fig:1}
\end{figure}

In Figure~\ref{fig:1}, we present the normalized oscillatory term of the electron density, $\delta R(r) = \pi\delta n(r)/(m k_F^\eta A)$, as a function of the dimensionless distance $k_F r$, as defined in Eq.~\eqref{eq:13}, for various values of number $\eta$. Our observations indicate that $\eta$, which characterizes the non-Coulomb nature of the impurity potential, influences only the amplitude of the oscillations while leaving their phase unchanged at sufficiently large distances from the perturbing impurity.

\begin{figure}
	\includegraphics[width=14cm,keepaspectratio]{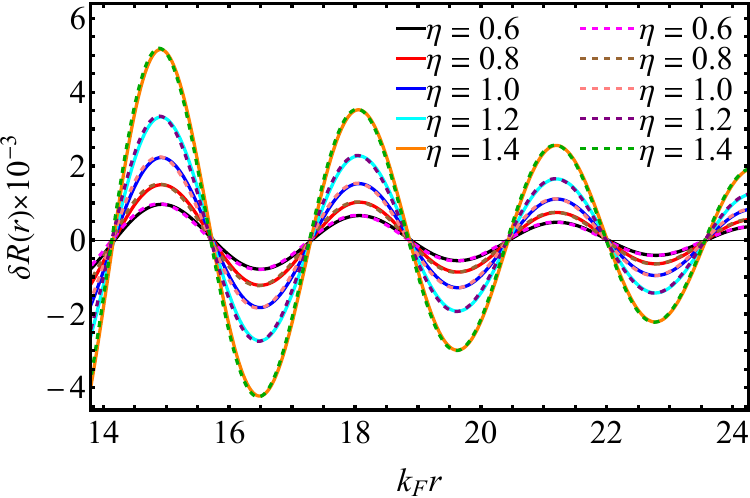}
	\caption{The normalized oscillatory term of the electron density $\delta R(r) = \pi\delta n(r)/(m k_F^\eta A)$, calculated with Eq.~\eqref{eq:13} (solid lines), and its asymptotic behavior approximated with Eq.~\eqref{eq:14} (dashed lines), as a function of the dimensionless distance $k_F r$ for two-dimensional non-interacting electron gas with different values of the number $\eta$, and $k_F=1$.}
	\label{fig:2}
\end{figure}
The asymptotic behavior of Eq.~\eqref{eq:13} at large-distances ($r \to \infty$) is approximated by
\begin{equation}
	\delta n(r) \sim -\frac{2 \, m \, k_F^\eta A}{[\Gamma (\frac{\eta}{2} )]^2 \sin (\frac{\pi\eta}{2})} \frac{\sin(2 k_F r)}{(2 k_F r)^2},
	\label{eq:14}
\end{equation}
which indicates that the parameter $\eta$ affects only the amplitude of oscillations at large distances and has no influence on their phase. The phase shift in Friedel oscillations is directly related to the scattering phase shift experienced by the electrons due to the impurity potential. According to quantum scattering theory, when a particle encounters a scattering potential its wavefunction undergoes a change in phase. In the Born approximation the phase shift is small. The expression for $n(r)$, in our model, is quite accurate for most values of the distance, except the vicinity of the impurity. Here the linear approximation is no more valid due to the fact that $n(r)$ becomes large enough. In this region, where small differences appear in the $r$ dependence of the normalized oscillatory term of electron density with $\eta$ in Figure~\ref{fig:1}, the validity of the model is questionable. However, working beyond the linear approximation, an influence of the $\eta$ parameter on the phase shift is possible.

We compare in Figure~\ref{fig:2} the results for the normalized oscillatory term of the electron density, $\delta R(r) = \pi\delta n(r)/(m k_F^\eta A)$, defined in Eq.~\eqref{eq:13} (see solid lines), with its large distance approximated form in Eq.~\eqref{eq:14} (see dashed lines), as a function of the dimensionless distance $k_F r$, for different values of the parameter $\eta$. We observe that the large-distance behavior of the oscillatory term of the electron density given by Eq.~\eqref{eq:14} approximates well the results obtained with the help of Eq.~\eqref{eq:13} at large distances measured from the perturbing potential, confirming thus its validity, even in the $\eta = 1$ limit.   

In the special $\eta = 1$ case, corresponding to the Coulomb-like potential, Eq.~\eqref{eq:14} reduces to
\begin{equation}
	\delta n(r) \sim -\frac{2 m k_F A}{\pi} \frac{\sin(2k_F r)}{(2k_F r)^2},
	\label{eq:15}
\end{equation}
which is consistent with the results of Refs.~\cite{Giuliani2003,Simion2005,Giuliani2005a} for a long-range (Coulomb-like) impurity potential.\\

\textit{Case (ii): Monolayer graphene}

Substituting Eq.~\eqref{eq:5} into Eq.~\eqref{eq:2}, we obtain
\begin{equation}
	\begin{split}
	n(r)&=D(E_F)\bigg \{ \int \frac{d^2 q}{(2\pi)^2}  e^{i \vec{q}\cdot\vec{r}}\, V(q)-\int \frac{d^2 q}{(2\pi)^2}  e^{i \vec{q}\cdot\vec{r}} \, V(q)\\
	&\times \theta(q-2k_F)\bigg [ \frac{1}{2} \sqrt{1-\Big (\frac{2k_F}{q}\Big )^2} +\frac{q}{4k_F} \arcsin(\frac{2k_F}{q}) -\frac{\pi q}{8k_F} \bigg ]\,\bigg \},
	\end{split}
	\label{eq:16}
\end{equation}
which, together with Eq.~\eqref{eq:3}, in polar coordinates, transforms to
\begin{equation}
	\begin{split}
		n(r)&=A D(E_F)\bigg \{\frac{1}{r^\eta}-\frac{2^{2-\eta} [\Gamma(1-\frac{\eta}{2})]^2 \sin(\frac{\pi \eta}{2})}{(2\pi)^2}\int_{0}^{\infty} \frac{dq \, q}{q^{2-\eta}} \int_{0}^{2\pi}d\phi \, e^{i q \, r\, \cos\phi} \\
		&\times \theta(q-2k_F) \bigg [ \frac{1}{2} \sqrt{1-\Big (\frac{2k_F}{q}\Big )^2} +\frac{q}{4k_F} \arcsin(\frac{2k_F}{q}) -\frac{\pi q}{8k_F} \bigg ]\,\bigg \},
	\end{split}
	\label{eq:17}
\end{equation}
that reduces to
\begin{equation}
	\begin{split}
		n(r)&=A D(E_F)\bigg \{\frac{1}{r^\eta}-\frac{[\Gamma(1-\frac{\eta}{2})]^2 \sin(\frac{\pi \eta}{2})}{2^{\eta}\,\pi}\int_{2k_F}^{\infty} dq \, q^{\eta-1} J_0(q \, r) \\
		&\times \bigg [\sqrt{1-\Big (\frac{2k_F}{q}\Big )^2} +\frac{q}{2k_F} \arcsin(\frac{2k_F}{q}) -\frac{\pi q}{4k_F} \bigg ]\,\bigg \}.
	\end{split}
	\label{eq:18}
\end{equation}
Introducing now the variable $x=q/2k_F$, we have
\begin{equation}
	\begin{split}
		n(r)&=A D(E_F)\bigg \{\frac{1}{r^\eta}-\frac{k_F^\eta \, [\Gamma(1-\frac{\eta}{2})]^2 \sin(\frac{\pi \eta}{2})}{\pi}  \int_{1}^{\infty} dx \, x^{\eta-1} J_0(2 k_F r x) \\
		&\times \bigg [\frac{\sqrt{x^2 - 1}}{x} + x \arcsin(\frac{1}{x}) -\frac{\pi}{2} x \bigg ]\,\bigg \} \equiv D(E_F)V(r)+\delta n(r).
	\end{split}
	\label{eq:19}
\end{equation}
Following similar steps to those in the case of the two-dimensional non-interacting electron gas, we find (details are given in Appendix~\ref{sec:A}):
\begin{equation}
	\delta n(r)\approx-A k_F^\eta D(E_F)\frac{1}{(k_F r)^\eta}\bigg [\frac{1}{2}\mathcal{F}_1(r)+(k_F \, r)^\eta \mathcal{N}(r)\bigg ],
	\label{eq:20}
\end{equation}
where
\begin{equation}
	\begin{split}
	\mathcal{F}_1(r)& ={}_1F_2\Big (-\frac{1}{2};1-\frac{\eta}{2},1-\frac{\eta}{2};-(k_F r)^2\Big )\\
	&+ {}_2F_3 \Big (\frac{1}{2},\frac{1}{2};\frac{3}{2},1-\frac{\eta}{2},1-\frac{\eta}{2};-(k_F r)^2\Big ),
	\end{split}
	\label{eq:21}
\end{equation}
and
\begin{equation}
	\mathcal{N}(r)=\frac{\eta^2 [\Gamma(-\frac{\eta}{2})]^3 \sin(\frac{\pi \eta}{2})}{16 \sqrt{\pi}\,\Gamma(\frac{3-\eta}{2})} \bigg [\mathcal{F}_2(r)+\frac{1-\eta}{1+\eta}\mathcal{F}_3(r) \bigg ],
	\label{eq:22}
\end{equation}
with
\begin{equation}
	\mathcal{F}_2 (r)={}_1F_2\Big (\frac{\eta - 1}{2};1,1+\frac{\eta}{2};-(k_F r)^2\Big ),
	\label{eq:23}
\end{equation}
and
\begin{equation}
		\mathcal{F}_3 (r) ={}_2F_3 \bigg ( \frac{\eta + 1}{2},\frac{\eta + 1}{2};1,1+\frac{\eta}{2},\frac{\eta + 3}{2};-(k_F r)^2 \bigg ).
	\label{eq:24}
\end{equation}

In the case of Coulomb-like potential ($\eta = 1$), Eq.~\eqref{eq:20} reduces to the form
\begin{equation}
	\delta n(r)\approx -\frac{A k_F D(E_F)}{2 k_F r} [\mathcal{R} (r)-\pi k_F r],
	\label{eq:25}
\end{equation}
with
\begin{equation}
	\mathcal{R} (r) = {}_1F_2\Big (-\frac{1}{2};\frac{1}{2},\frac{1}{2};-(k_F r)^2 \Big ) + {}_2F_3 \Big ( \frac{1}{2},\frac{1}{2};\frac{3}{2},\frac{1}{2},\frac{1}{2};-(k_F r)^2 \Big ).
	\label{eq:26}
\end{equation}
The asymptotic behavior of Eq.~\eqref{eq:25} at large-distances ($r \to \infty$) is approximated by
\begin{equation}
	\delta n(r)\sim -2A k_F D(E_F)\frac{\cos(2k_F r)}{(2k_F r)^3}.
	\label{eq:27}
\end{equation}
This result is in qualitative agreement with those of Ref.~\cite{Wunsch2006,Cheianov2006}.

\begin{figure}
	\includegraphics[width=14cm,keepaspectratio]{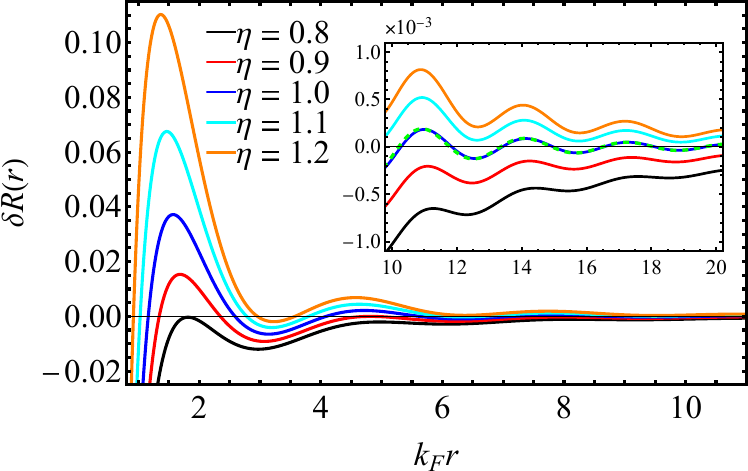}
	\caption{The normalized oscillatory term of charge carrier density $\delta R(r) = \delta n(r)/[A k_F^\eta D(E_F)]$ as a function of the dimensionless distance $k_F r$ for monolayer graphene with different values of the number $\eta$ and $k_F=1$. The inset of figure shows the results for a larger distance where the green dashed line corresponds to the large-distance asymptotic behavior of $\delta R(r)$ for $\eta = 1$.}
	\label{fig:3}
\end{figure}

In Figure~\ref{fig:3}, we plot the normalized oscillatory term of charge carrier density, given in Eq.~\eqref{eq:20}, $\delta R(r) = \delta n (r)/[A k_F^\eta D(E_F)]$, as a function of dimensionless distance $k_F r$ for different values of the number $\eta$. The inset of Figure~\ref{fig:3} illustrates $\delta R(r) = \delta n (r)/[A k_F^\eta D(E_F)]$ as a function of dimensionless distance $k_F r$ for the same values of $\eta$ at higher distances with the large-distance asymptotic approximation of $\delta R(r)$ for a fixed $\eta = 1$ value (dashed line), given by Eq.~\eqref{eq:27}.
We see that the parameter $\eta$ strongly affects the amplitude of charge carrier density oscillations as a function of $k_F r$ and the large-distance asymptotic approximation of $\delta n (r)$ in the $\eta = 1$ case matches well the finding calculated with relation~\eqref{eq:20} confirming thus its validity. In addition, the parameter $\eta$ shows approximately no impact on the phase of Friedel oscillations at large distances from the perturbing impurity potential. Small deviations that may be observed in the phase of charge oscillations can be attributed to the approximations used in deriving $\delta n(r)$ in Eq.~\eqref{eq:20}, discussed in Appendix~\ref{sec:A}. Moreover, the overall form of the oscillations, such as their amplitude modulation and asymptotic decay, are also influenced by these approximations. Here also, for small distances from impurities, the validity of the model is questionable, in higher-order approximations an influence of the $\eta$ parameter on the phase shift may become significant.

\section{Discussions}
\label{sec:III}
In this paper, we analyzed Friedel oscillations in two-dimensional systems under a non-Coulomb impurity potential described by the phenomenological $r^{-\eta}$ form. Friedel oscillations are essential for understanding the electronic properties of materials in the presence of impurities. They play an important role in determining the scattering of electrons, which affects the electrical conductivity and other important transport properties. In nanoscale devices the impurities can have a significant impact, and understanding Friedel oscillations is vital to optimizing the performance of these devices. On the other hand, the analytical form of the impurity potential is important, for the non-Coulombian case being both theoretical and experimental approaches (see Refs.~\cite{Katsnelson2006,Gabovich2019,Vangara2019,Zhang2022,Defenu2024,Radzihovsky2024}). The analytical results obtained in this paper generalize the results obtained for the classical two-dimensional electronic systems, as well as for graphene, considering the potential of the form $r^{-\eta}$. The previous results, for the Coulombian-like potential and for large distances, are reobtained by taking the limit $\eta =1$. In addition, for large distances and in the case of non-Coulomb interactions, the behavior of the Friedel oscillations was determined. The non-Coulomb character of the interaction affects, as expected, the amplitude of the oscillations.
Note that the problem of analytical form of the potential is important. In a strictly two-dimensional system, the potential takes a logarithmic form, as this being a solution of the Poisson equation~\cite{Kraut2007}. This arises from the different dimensionality of space, which affects the divergence of the electric field. However, in two-dimensional systems embedded in a three-dimensional space, the Coulomb potential is more suitable to be used rather than the logarithmic one (see Refs.~\cite{Giuliani2005a,Katsnelson2020,Hwang2007}). Furthermore, considering the two-dimensional system as being obtained from a Coulomb three-dimensional layered system whose thickness tends to zero~\cite{Roesner2015}, the classical $1/|q|$ result is recovered.
Another important case is the screened Coulomb interaction, which could be the subject of a future research work. In this case, the potential has a short-range character, and there may be similarities with the cases discussed here, for $\eta>1$, when the potential also has a shorter range behavior~\cite{Min2012,Chen2015,Thakur2017}.

\appendix
\section{}
\label{sec:A}
In order to obtain Eq.~\eqref{eq:20} from Eq.~\eqref{eq:19}, we use the expression for $I_0$ given in Eq.~\eqref{eq:10} along with the following integrals:
\begin{equation}
	\begin{split}
		I_1 &=\int_{1}^{\infty} dx \, x^{\eta} J_0(2 k_F r x) \arcsin(\frac{1}{x})\\
		&=-\frac{\pi}{2(1+\eta)} \, {}_1F_2\bigg ( \frac{\eta +1}{2};1,\frac{\eta + 3}{2};-(k_F r)^2 \bigg)\\
		&+\frac{\Gamma(\frac{\eta}{2})}{2\,\Gamma(1-\frac{\eta}{2})} (k_F r)^{-\eta} \, {}_2F_3 \bigg ( \frac{1}{2},\frac{1}{2};\frac{3}{2},1-\frac{\eta}{2},1-\frac{\eta}{2};-(k_F r)^2 \bigg )\\
		&-\frac{\sqrt{\pi}\, \Gamma(-\frac{\eta}{2})}{(1+\eta)^2 \,\Gamma(-\frac{\eta +1}{2})} \, {}_2F_3 \bigg ( \frac{\eta+1}{2},\frac{\eta+1}{2};1,1+\frac{\eta}{2},\frac{\eta + 3}{2};-(k_F r)^2 \bigg ),
	\end{split}
	\label{eq:A1}
\end{equation}
valid for $\eta< \frac{3}{2}$ and
\begin{equation}
	\begin{split}
		I_2 &= \int_{1}^{\infty} dx \,\frac{\pi}{2} x^{\eta} J_0(2 k_F r x) \approx \int_{1}^{q_c/2k_F} dx \,\frac{\pi}{2} x^{\eta} J_0(2 k_F r x)\\
		&= \int_{0}^{q_c/2k_F} dx \,\frac{\pi}{2} x^{\eta} J_0(2 k_F r x)-\int_{0}^{1} dx \,\frac{\pi}{2} x^{\eta} J_0(2 k_F r x).
	\end{split}
	\label{eq:A2}
\end{equation}
Here, an upper limit of integration is introduced to make the integral analytically tractable, especially for large upper-limit values. The first term of integral~\eqref{eq:A2} exhibits a rapidly oscillating behavior as a function of distance. The frequency of these oscillations increases with the momentum cutoff $q_c$ and near $\eta \approx 1$ the positive and negative contributions of the oscillations approximately cancel each other out. Moreover, in contrast to the other terms, the contribution of this integral to charge carrier density deviations at large distances from the perturbing impurity may be negligible. Additionally, the highly oscillating behavior induced by the momentum cutoff $q_c$ appears unphysical, as the resulting oscillations occur to be meaningfully considered~\cite{Bacsi2010}.

Using now the identity~\cite{Luke1969}: 
\begin{equation}
	\int_{0}^{z} dx \, x^\mu J_\nu (x) = \frac{z^{\mu + \nu +1}}{2^\nu \,(\mu + \nu + 1)\,\Gamma(\nu + 1)}\,{}_1F_2 \bigg ( \frac{\mu + \nu + 1}{2};\frac{\mu + \nu +3}{2}, \nu +1;-\frac{z^2}{4} \bigg ),
	\label{eq:A3}
\end{equation}
valid for $\Re (\mu + \nu) > -1$, the second term of $I_2$ is calculated as
\begin{equation}
	I_2 \approx -\frac{\pi}{2(1+\eta)} {}_1F_2\Big ( \frac{\eta + 1}{2}; \frac{\eta + 3}{2}, 1; -(k_F r)^2 \Big ).
\label{eq:A4}
\end{equation}
Using these results, together with the first line from Eq.~\eqref{eq:11}, we obtain the result given in Eq.~\eqref{eq:20}.

\begin{acknowledgements}
The authors would like to thank Dr.~Doru~Sticleț for valuable discussions. This work was supported through the ``Nucleu'' Program within the National Research Development and Innovation Plan 2022–2027, Romania, carried out with the support of MEC, project no. 27N/03.01.2023, component project code PN 23 24 01 04 and also received financial support from CNCS/CCCDI-UEFISCDI, under project number PN-IV-P1-PCE-2023-0987.
\end{acknowledgements}

\bibliographystyle{apsrev4-2}  
\bibliography{ReferencesFriedel}

\begin{thebibliography}{62}%
\makeatletter
\providecommand \@ifxundefined [1]{%
 \@ifx{#1\undefined}
}%
\providecommand \@ifnum [1]{%
 \ifnum #1\expandafter \@firstoftwo
 \else \expandafter \@secondoftwo
 \fi
}%
\providecommand \@ifx [1]{%
 \ifx #1\expandafter \@firstoftwo
 \else \expandafter \@secondoftwo
 \fi
}%
\providecommand \natexlab [1]{#1}%
\providecommand \enquote  [1]{``#1''}%
\providecommand \bibnamefont  [1]{#1}%
\providecommand \bibfnamefont [1]{#1}%
\providecommand \citenamefont [1]{#1}%
\providecommand \href@noop [0]{\@secondoftwo}%
\providecommand \href [0]{\begingroup \@sanitize@url \@href}%
\providecommand \@href[1]{\@@startlink{#1}\@@href}%
\providecommand \@@href[1]{\endgroup#1\@@endlink}%
\providecommand \@sanitize@url [0]{\catcode `\\12\catcode `\$12\catcode
  `\&12\catcode `\#12\catcode `\^12\catcode `\_12\catcode `\%12\relax}%
\providecommand \@@startlink[1]{}%
\providecommand \@@endlink[0]{}%
\providecommand \url  [0]{\begingroup\@sanitize@url \@url }%
\providecommand \@url [1]{\endgroup\@href {#1}{\urlprefix }}%
\providecommand \urlprefix  [0]{URL }%
\providecommand \Eprint [0]{\href }%
\providecommand \doibase [0]{https://doi.org/}%
\providecommand \selectlanguage [0]{\@gobble}%
\providecommand \bibinfo  [0]{\@secondoftwo}%
\providecommand \bibfield  [0]{\@secondoftwo}%
\providecommand \translation [1]{[#1]}%
\providecommand \BibitemOpen [0]{}%
\providecommand \bibitemStop [0]{}%
\providecommand \bibitemNoStop [0]{.\EOS\space}%
\providecommand \EOS [0]{\spacefactor3000\relax}%
\providecommand \BibitemShut  [1]{\csname bibitem#1\endcsname}%
\let\auto@bib@innerbib\@empty
\bibitem [{\citenamefont {Friedel}(1952)}]{Friedel1952}%
  \BibitemOpen
  \bibfield  {author} {\bibinfo {author} {\bibfnamefont {J.}~\bibnamefont
  {Friedel}},\ }\href {https://doi.org/10.1080/14786440208561086} {\bibfield
  {journal} {\bibinfo  {journal} {Lond. Edinburgh Dublin Philos. Mag. J. Sci.}\
  }\textbf {\bibinfo {volume} {43}},\ \bibinfo {pages} {153} (\bibinfo {year}
  {1952})}\BibitemShut {NoStop}%
\bibitem [{\citenamefont {Villain}\ \emph {et~al.}(2015)\citenamefont
  {Villain}, \citenamefont {Lavagna},\ and\ \citenamefont
  {Bruno}}]{Villain2015}%
  \BibitemOpen
  \bibfield  {author} {\bibinfo {author} {\bibfnamefont {J.}~\bibnamefont
  {Villain}}, \bibinfo {author} {\bibfnamefont {M.}~\bibnamefont {Lavagna}},\
  and\ \bibinfo {author} {\bibfnamefont {P.}~\bibnamefont {Bruno}},\ }\href
  {https://doi.org/10.1016/j.crhy.2015.12.010} {\bibfield  {journal} {\bibinfo
  {journal} {Comptes Rendus. Physique}\ }\textbf {\bibinfo {volume} {17}},\
  \bibinfo {pages} {276} (\bibinfo {year} {2015})}\BibitemShut {NoStop}%
\bibitem [{\citenamefont {Giuliani}\ and\ \citenamefont
  {Vignale}(2005)}]{Giuliani2005a}%
  \BibitemOpen
  \bibfield  {author} {\bibinfo {author} {\bibfnamefont {G.}~\bibnamefont
  {Giuliani}}\ and\ \bibinfo {author} {\bibfnamefont {G.}~\bibnamefont
  {Vignale}},\ }\href {https://doi.org/10.1017/cbo9780511619915} {\emph
  {\bibinfo {title} {Quantum Theory of the Electron Liquid}}}\ (\bibinfo
  {publisher} {Cambridge University Press},\ \bibinfo {year}
  {2005})\BibitemShut {NoStop}%
\bibitem [{\citenamefont {Wunsch}\ \emph {et~al.}(2006)\citenamefont {Wunsch},
  \citenamefont {Stauber}, \citenamefont {Sols},\ and\ \citenamefont
  {Guinea}}]{Wunsch2006}%
  \BibitemOpen
  \bibfield  {author} {\bibinfo {author} {\bibfnamefont {B.}~\bibnamefont
  {Wunsch}}, \bibinfo {author} {\bibfnamefont {T.}~\bibnamefont {Stauber}},
  \bibinfo {author} {\bibfnamefont {F.}~\bibnamefont {Sols}},\ and\ \bibinfo
  {author} {\bibfnamefont {F.}~\bibnamefont {Guinea}},\ }\href
  {https://doi.org/10.1088/1367-2630/8/12/318} {\bibfield  {journal} {\bibinfo
  {journal} {New J. Phys.}\ }\textbf {\bibinfo {volume} {8}},\ \bibinfo {pages}
  {318} (\bibinfo {year} {2006})}\BibitemShut {NoStop}%
\bibitem [{\citenamefont {Cheianov}\ and\ \citenamefont
  {Fal'ko}(2006)}]{Cheianov2006}%
  \BibitemOpen
  \bibfield  {author} {\bibinfo {author} {\bibfnamefont {V.~V.}\ \bibnamefont
  {Cheianov}}\ and\ \bibinfo {author} {\bibfnamefont {V.~I.}\ \bibnamefont
  {Fal'ko}},\ }\href {https://doi.org/10.1103/PhysRevLett.97.226801} {\bibfield
   {journal} {\bibinfo  {journal} {Phys. Rev. Lett.}\ }\textbf {\bibinfo
  {volume} {97}},\ \bibinfo {pages} {226801} (\bibinfo {year}
  {2006})}\BibitemShut {NoStop}%
\bibitem [{\citenamefont {Pyatkovskiy}(2008)}]{Pyatkovskiy2008}%
  \BibitemOpen
  \bibfield  {author} {\bibinfo {author} {\bibfnamefont {P.~K.}\ \bibnamefont
  {Pyatkovskiy}},\ }\href {https://doi.org/10.1088/0953-8984/21/2/025506}
  {\bibfield  {journal} {\bibinfo  {journal} {J. Phys.: Condens. Matter}\
  }\textbf {\bibinfo {volume} {21}},\ \bibinfo {pages} {025506} (\bibinfo
  {year} {2008})}\BibitemShut {NoStop}%
\bibitem [{\citenamefont {B\'acsi}\ and\ \citenamefont
  {Virosztek}(2010)}]{Bacsi2010}%
  \BibitemOpen
  \bibfield  {author} {\bibinfo {author} {\bibfnamefont {A.}~\bibnamefont
  {B\'acsi}}\ and\ \bibinfo {author} {\bibfnamefont {A.}~\bibnamefont
  {Virosztek}},\ }\href {https://doi.org/10.1103/PhysRevB.82.193405} {\bibfield
   {journal} {\bibinfo  {journal} {Phys. Rev. B}\ }\textbf {\bibinfo {volume}
  {82}},\ \bibinfo {pages} {193405} (\bibinfo {year} {2010})}\BibitemShut
  {NoStop}%
\bibitem [{\citenamefont {Thakur}\ \emph {et~al.}(2017)\citenamefont {Thakur},
  \citenamefont {Sachdeva},\ and\ \citenamefont {Agarwal}}]{Thakur2017}%
  \BibitemOpen
  \bibfield  {author} {\bibinfo {author} {\bibfnamefont {A.}~\bibnamefont
  {Thakur}}, \bibinfo {author} {\bibfnamefont {R.}~\bibnamefont {Sachdeva}},\
  and\ \bibinfo {author} {\bibfnamefont {A.}~\bibnamefont {Agarwal}},\ }\href
  {https://doi.org/10.1088/1361-648X/aa57bd} {\bibfield  {journal} {\bibinfo
  {journal} {J. Phys.: Condens. Matter}\ }\textbf {\bibinfo {volume} {29}},\
  \bibinfo {pages} {105701} (\bibinfo {year} {2017})}\BibitemShut {NoStop}%
\bibitem [{\citenamefont {Katsnelson}(2020)}]{Katsnelson2020}%
  \BibitemOpen
  \bibfield  {author} {\bibinfo {author} {\bibfnamefont {M.~I.}\ \bibnamefont
  {Katsnelson}},\ }\href {https://doi.org/10.1017/9781108617567} {\emph
  {\bibinfo {title} {The Physics of Graphene}}}\ (\bibinfo  {publisher}
  {Cambridge University Press},\ \bibinfo {year} {2020})\BibitemShut {NoStop}%
\bibitem [{\citenamefont {Elsayed}\ \emph {et~al.}(2023)\citenamefont
  {Elsayed}, \citenamefont {Kim}, \citenamefont {Vanegas},\ and\ \citenamefont
  {Kotov}}]{Elsayed2023}%
  \BibitemOpen
  \bibfield  {author} {\bibinfo {author} {\bibfnamefont {M.~M.}\ \bibnamefont
  {Elsayed}}, \bibinfo {author} {\bibfnamefont {S.~W.}\ \bibnamefont {Kim}},
  \bibinfo {author} {\bibfnamefont {J.~M.}\ \bibnamefont {Vanegas}},\ and\
  \bibinfo {author} {\bibfnamefont {V.~N.}\ \bibnamefont {Kotov}},\ }\href
  {https://doi.org/10.1103/PhysRevB.108.245414} {\bibfield  {journal} {\bibinfo
   {journal} {Phys. Rev. B}\ }\textbf {\bibinfo {volume} {108}},\ \bibinfo
  {pages} {245414} (\bibinfo {year} {2023})}\BibitemShut {NoStop}%
\bibitem [{\citenamefont {Lin}\ and\ \citenamefont {Chiu}(2024)}]{Lin2024}%
  \BibitemOpen
  \bibfield  {author} {\bibinfo {author} {\bibfnamefont {C.-Y.}\ \bibnamefont
  {Lin}}\ and\ \bibinfo {author} {\bibfnamefont {C.-W.}\ \bibnamefont {Chiu}},\
  }\href {https://doi.org/10.1038/s41598-024-63738-w} {\bibfield  {journal}
  {\bibinfo  {journal} {Sci. Rep.}\ }\textbf {\bibinfo {volume} {14}},\
  \bibinfo {pages} {13792} (\bibinfo {year} {2024})}\BibitemShut {NoStop}%
\bibitem [{\citenamefont {Giuliani}\ \emph {et~al.}(2005)\citenamefont
  {Giuliani}, \citenamefont {Vignale},\ and\ \citenamefont
  {Datta}}]{Giuliani2005}%
  \BibitemOpen
  \bibfield  {author} {\bibinfo {author} {\bibfnamefont {G.~F.}\ \bibnamefont
  {Giuliani}}, \bibinfo {author} {\bibfnamefont {G.}~\bibnamefont {Vignale}},\
  and\ \bibinfo {author} {\bibfnamefont {T.}~\bibnamefont {Datta}},\ }\href
  {https://doi.org/10.1103/PhysRevB.72.033411} {\bibfield  {journal} {\bibinfo
  {journal} {Phys. Rev. B}\ }\textbf {\bibinfo {volume} {72}},\ \bibinfo
  {pages} {033411} (\bibinfo {year} {2005})}\BibitemShut {NoStop}%
\bibitem [{\citenamefont {Grosu}\ and\ \citenamefont
  {Tugulan}(2008)}]{Grosu2008}%
  \BibitemOpen
  \bibfield  {author} {\bibinfo {author} {\bibfnamefont {I.}~\bibnamefont
  {Grosu}}\ and\ \bibinfo {author} {\bibfnamefont {L.}~\bibnamefont
  {Tugulan}},\ }\href {https://doi.org/10.1007/s10948-007-0300-1} {\bibfield
  {journal} {\bibinfo  {journal} {J. Supercond. Novel Magn.}\ }\textbf
  {\bibinfo {volume} {21}},\ \bibinfo {pages} {65} (\bibinfo {year}
  {2008})}\BibitemShut {NoStop}%
\bibitem [{\citenamefont {Máthé}\ and\ \citenamefont
  {Grosu}(2023)}]{Mathe2023}%
  \BibitemOpen
  \bibfield  {author} {\bibinfo {author} {\bibfnamefont {L.}~\bibnamefont
  {Máthé}}\ and\ \bibinfo {author} {\bibfnamefont {I.}~\bibnamefont
  {Grosu}},\ }\href {https://doi.org/10.24193/subbphys.2023.05} {\bibfield
  {journal} {\bibinfo  {journal} {Stud. Univ. Babes-Bolyai, Physica}\ }\textbf
  {\bibinfo {volume} {68 (LXVIII)}},\ \bibinfo {pages} {49} (\bibinfo {year}
  {2023})}\BibitemShut {NoStop}%
\bibitem [{\citenamefont {Tugulan}(2008)}]{Tugulan2008}%
  \BibitemOpen
  \bibfield  {author} {\bibinfo {author} {\bibfnamefont {L.}~\bibnamefont
  {Tugulan}},\ }\href@noop {} {\bibfield  {journal} {\bibinfo  {journal} {Stud.
  Univ. Babes-Bolyai, Physica}\ }\textbf {\bibinfo {volume} {1 (LIII)}},\
  \bibinfo {pages} {33} (\bibinfo {year} {2008})}\BibitemShut {NoStop}%
\bibitem [{\citenamefont {Grosu}\ and\ \citenamefont
  {Tugulan}(2007)}]{Grosu2007}%
  \BibitemOpen
  \bibfield  {author} {\bibinfo {author} {\bibfnamefont {I.}~\bibnamefont
  {Grosu}}\ and\ \bibinfo {author} {\bibfnamefont {L.}~\bibnamefont
  {Tugulan}},\ }\href@noop {} {\bibfield  {journal} {\bibinfo  {journal} {Stud.
  Univ. Babes-Bolyai, Physica}\ }\textbf {\bibinfo {volume} {2 (LII)}},\
  \bibinfo {pages} {79} (\bibinfo {year} {2007})}\BibitemShut {NoStop}%
\bibitem [{\citenamefont {Fabrizio}\ and\ \citenamefont
  {Gogolin}(1995)}]{Fabrizio1995}%
  \BibitemOpen
  \bibfield  {author} {\bibinfo {author} {\bibfnamefont {M.}~\bibnamefont
  {Fabrizio}}\ and\ \bibinfo {author} {\bibfnamefont {A.~O.}\ \bibnamefont
  {Gogolin}},\ }\href {https://doi.org/10.1103/PhysRevB.51.17827} {\bibfield
  {journal} {\bibinfo  {journal} {Phys. Rev. B}\ }\textbf {\bibinfo {volume}
  {51}},\ \bibinfo {pages} {17827} (\bibinfo {year} {1995})}\BibitemShut
  {NoStop}%
\bibitem [{\citenamefont {Gorczyca}\ \emph {et~al.}(2007)\citenamefont
  {Gorczyca}, \citenamefont {Ma\ifmmode~\acute{s}\else \'{s}\fi{}ka},\ and\
  \citenamefont {Mierzejewski}}]{Gorczyca2007}%
  \BibitemOpen
  \bibfield  {author} {\bibinfo {author} {\bibfnamefont {A.}~\bibnamefont
  {Gorczyca}}, \bibinfo {author} {\bibfnamefont {M.~M.}\ \bibnamefont
  {Ma\ifmmode~\acute{s}\else \'{s}\fi{}ka}},\ and\ \bibinfo {author}
  {\bibfnamefont {M.}~\bibnamefont {Mierzejewski}},\ }\href
  {https://doi.org/10.1103/PhysRevB.76.165419} {\bibfield  {journal} {\bibinfo
  {journal} {Phys. Rev. B}\ }\textbf {\bibinfo {volume} {76}},\ \bibinfo
  {pages} {165419} (\bibinfo {year} {2007})}\BibitemShut {NoStop}%
\bibitem [{\citenamefont {Gorczyca}\ \emph {et~al.}(2009)\citenamefont
  {Gorczyca}, \citenamefont {Maśka},\ and\ \citenamefont
  {Mierzejewski}}]{Gorczyca2009}%
  \BibitemOpen
  \bibfield  {author} {\bibinfo {author} {\bibfnamefont {A.}~\bibnamefont
  {Gorczyca}}, \bibinfo {author} {\bibfnamefont {M.~M.}\ \bibnamefont
  {Maśka}},\ and\ \bibinfo {author} {\bibfnamefont {M.}~\bibnamefont
  {Mierzejewski}},\ }\href {https://doi.org/10.1002/pssb.200881557} {\bibfield
  {journal} {\bibinfo  {journal} {Phys. Status Solidi B}\ }\textbf {\bibinfo
  {volume} {246}},\ \bibinfo {pages} {989} (\bibinfo {year}
  {2009})}\BibitemShut {NoStop}%
\bibitem [{\citenamefont {S\"offing}\ \emph {et~al.}(2009)\citenamefont
  {S\"offing}, \citenamefont {Bortz}, \citenamefont {Schneider}, \citenamefont
  {Struck}, \citenamefont {Fleischhauer},\ and\ \citenamefont
  {Eggert}}]{Soffing2009}%
  \BibitemOpen
  \bibfield  {author} {\bibinfo {author} {\bibfnamefont {S.~A.}\ \bibnamefont
  {S\"offing}}, \bibinfo {author} {\bibfnamefont {M.}~\bibnamefont {Bortz}},
  \bibinfo {author} {\bibfnamefont {I.}~\bibnamefont {Schneider}}, \bibinfo
  {author} {\bibfnamefont {A.}~\bibnamefont {Struck}}, \bibinfo {author}
  {\bibfnamefont {M.}~\bibnamefont {Fleischhauer}},\ and\ \bibinfo {author}
  {\bibfnamefont {S.}~\bibnamefont {Eggert}},\ }\href
  {https://doi.org/10.1103/PhysRevB.79.195114} {\bibfield  {journal} {\bibinfo
  {journal} {Phys. Rev. B}\ }\textbf {\bibinfo {volume} {79}},\ \bibinfo
  {pages} {195114} (\bibinfo {year} {2009})}\BibitemShut {NoStop}%
\bibitem [{\citenamefont {Traverso~Ziani}\ \emph {et~al.}(2012)\citenamefont
  {Traverso~Ziani}, \citenamefont {Cavaliere},\ and\ \citenamefont
  {Sassetti}}]{Ziani2012}%
  \BibitemOpen
  \bibfield  {author} {\bibinfo {author} {\bibfnamefont {N.}~\bibnamefont
  {Traverso~Ziani}}, \bibinfo {author} {\bibfnamefont {F.}~\bibnamefont
  {Cavaliere}},\ and\ \bibinfo {author} {\bibfnamefont {M.}~\bibnamefont
  {Sassetti}},\ }\href {https://doi.org/10.1103/PhysRevB.86.125451} {\bibfield
  {journal} {\bibinfo  {journal} {Phys. Rev. B}\ }\textbf {\bibinfo {volume}
  {86}},\ \bibinfo {pages} {125451} (\bibinfo {year} {2012})}\BibitemShut
  {NoStop}%
\bibitem [{\citenamefont {Gambetta}\ \emph {et~al.}(2015)\citenamefont
  {Gambetta}, \citenamefont {Ziani}, \citenamefont {Barbarino}, \citenamefont
  {Cavaliere},\ and\ \citenamefont {Sassetti}}]{Gambetta2015}%
  \BibitemOpen
  \bibfield  {author} {\bibinfo {author} {\bibfnamefont {F.~M.}\ \bibnamefont
  {Gambetta}}, \bibinfo {author} {\bibfnamefont {N.~T.}\ \bibnamefont {Ziani}},
  \bibinfo {author} {\bibfnamefont {S.}~\bibnamefont {Barbarino}}, \bibinfo
  {author} {\bibfnamefont {F.}~\bibnamefont {Cavaliere}},\ and\ \bibinfo
  {author} {\bibfnamefont {M.}~\bibnamefont {Sassetti}},\ }\href
  {https://doi.org/10.1103/PhysRevB.91.235421} {\bibfield  {journal} {\bibinfo
  {journal} {Phys. Rev. B}\ }\textbf {\bibinfo {volume} {91}},\ \bibinfo
  {pages} {235421} (\bibinfo {year} {2015})}\BibitemShut {NoStop}%
\bibitem [{\citenamefont {Kyl\"anp\"a\"a}\ \emph {et~al.}(2016)\citenamefont
  {Kyl\"anp\"a\"a}, \citenamefont {Cavaliere}, \citenamefont {Ziani},
  \citenamefont {Sassetti},\ and\ \citenamefont {R\"as\"anen}}]{Kylanpaa2016}%
  \BibitemOpen
  \bibfield  {author} {\bibinfo {author} {\bibfnamefont {I.}~\bibnamefont
  {Kyl\"anp\"a\"a}}, \bibinfo {author} {\bibfnamefont {F.}~\bibnamefont
  {Cavaliere}}, \bibinfo {author} {\bibfnamefont {N.~T.}\ \bibnamefont
  {Ziani}}, \bibinfo {author} {\bibfnamefont {M.}~\bibnamefont {Sassetti}},\
  and\ \bibinfo {author} {\bibfnamefont {E.}~\bibnamefont {R\"as\"anen}},\
  }\href {https://doi.org/10.1103/PhysRevB.94.115417} {\bibfield  {journal}
  {\bibinfo  {journal} {Phys. Rev. B}\ }\textbf {\bibinfo {volume} {94}},\
  \bibinfo {pages} {115417} (\bibinfo {year} {2016})}\BibitemShut {NoStop}%
\bibitem [{\citenamefont {Ziani}\ \emph {et~al.}(2021)\citenamefont {Ziani},
  \citenamefont {Cavaliere}, \citenamefont {Becerra},\ and\ \citenamefont
  {Sassetti}}]{Ziani2021}%
  \BibitemOpen
  \bibfield  {author} {\bibinfo {author} {\bibfnamefont {N.~T.}\ \bibnamefont
  {Ziani}}, \bibinfo {author} {\bibfnamefont {F.}~\bibnamefont {Cavaliere}},
  \bibinfo {author} {\bibfnamefont {K.~G.}\ \bibnamefont {Becerra}},\ and\
  \bibinfo {author} {\bibfnamefont {M.}~\bibnamefont {Sassetti}},\ }\bibfield
  {journal} {\bibinfo  {journal} {Crystals}\ }\textbf {\bibinfo {volume}
  {11}},\ \href {https://doi.org/10.3390/cryst11010020} {10.3390/cryst11010020}
  (\bibinfo {year} {2021})\BibitemShut {NoStop}%
\bibitem [{\citenamefont {Machida}\ and\ \citenamefont
  {Koyama}(2003)}]{Machida2003}%
  \BibitemOpen
  \bibfield  {author} {\bibinfo {author} {\bibfnamefont {M.}~\bibnamefont
  {Machida}}\ and\ \bibinfo {author} {\bibfnamefont {T.}~\bibnamefont
  {Koyama}},\ }\href {https://doi.org/10.1103/PhysRevLett.90.077003} {\bibfield
   {journal} {\bibinfo  {journal} {Phys. Rev. Lett.}\ }\textbf {\bibinfo
  {volume} {90}},\ \bibinfo {pages} {077003} (\bibinfo {year}
  {2003})}\BibitemShut {NoStop}%
\bibitem [{\citenamefont {Lauke}\ \emph {et~al.}(2018)\citenamefont {Lauke},
  \citenamefont {Scheurer}, \citenamefont {Poenicke},\ and\ \citenamefont
  {Schmalian}}]{Lauke2018}%
  \BibitemOpen
  \bibfield  {author} {\bibinfo {author} {\bibfnamefont {L.}~\bibnamefont
  {Lauke}}, \bibinfo {author} {\bibfnamefont {M.~S.}\ \bibnamefont {Scheurer}},
  \bibinfo {author} {\bibfnamefont {A.}~\bibnamefont {Poenicke}},\ and\
  \bibinfo {author} {\bibfnamefont {J.}~\bibnamefont {Schmalian}},\ }\href
  {https://doi.org/10.1103/PhysRevB.98.134502} {\bibfield  {journal} {\bibinfo
  {journal} {Phys. Rev. B}\ }\textbf {\bibinfo {volume} {98}},\ \bibinfo
  {pages} {134502} (\bibinfo {year} {2018})}\BibitemShut {NoStop}%
\bibitem [{\citenamefont {Stosiek}\ \emph {et~al.}(2022)\citenamefont
  {Stosiek}, \citenamefont {Baretzky}, \citenamefont {Balashov}, \citenamefont
  {Evers},\ and\ \citenamefont {Wulfhekel}}]{Stosiek2022}%
  \BibitemOpen
  \bibfield  {author} {\bibinfo {author} {\bibfnamefont {M.}~\bibnamefont
  {Stosiek}}, \bibinfo {author} {\bibfnamefont {C.}~\bibnamefont {Baretzky}},
  \bibinfo {author} {\bibfnamefont {T.}~\bibnamefont {Balashov}}, \bibinfo
  {author} {\bibfnamefont {F.}~\bibnamefont {Evers}},\ and\ \bibinfo {author}
  {\bibfnamefont {W.}~\bibnamefont {Wulfhekel}},\ }\href
  {https://doi.org/10.1103/PhysRevB.105.L140504} {\bibfield  {journal}
  {\bibinfo  {journal} {Phys. Rev. B}\ }\textbf {\bibinfo {volume} {105}},\
  \bibinfo {pages} {L140504} (\bibinfo {year} {2022})}\BibitemShut {NoStop}%
\bibitem [{\citenamefont {D\'ora}\ \emph {et~al.}(2021)\citenamefont {D\'ora},
  \citenamefont {Sticlet},\ and\ \citenamefont {Moca}}]{Dora2021}%
  \BibitemOpen
  \bibfield  {author} {\bibinfo {author} {\bibfnamefont {B.}~\bibnamefont
  {D\'ora}}, \bibinfo {author} {\bibfnamefont {D.}~\bibnamefont {Sticlet}},\
  and\ \bibinfo {author} {\bibfnamefont {C.~P.}\ \bibnamefont {Moca}},\ }\href
  {https://doi.org/10.1103/PhysRevB.104.125113} {\bibfield  {journal} {\bibinfo
   {journal} {Phys. Rev. B}\ }\textbf {\bibinfo {volume} {104}},\ \bibinfo
  {pages} {125113} (\bibinfo {year} {2021})}\BibitemShut {NoStop}%
\bibitem [{\citenamefont {Sticlet}\ \emph {et~al.}(2024)\citenamefont
  {Sticlet}, \citenamefont {Dóra},\ and\ \citenamefont {Moca}}]{Sticlet2024}%
  \BibitemOpen
  \bibfield  {author} {\bibinfo {author} {\bibfnamefont {D.}~\bibnamefont
  {Sticlet}}, \bibinfo {author} {\bibfnamefont {B.}~\bibnamefont {Dóra}},\
  and\ \bibinfo {author} {\bibfnamefont {C.~P.}\ \bibnamefont {Moca}},\ }in\
  \href {https://doi.org/10.1063/5.0215417} {\emph {\bibinfo {booktitle} {AIP
  Conf. Proc.}}},\ Vol.\ \bibinfo {volume} {3181}\ (\bibinfo {year} {2024})\
  p.\ \bibinfo {pages} {030001}\BibitemShut {NoStop}%
\bibitem [{\citenamefont {Giuliani}\ and\ \citenamefont
  {Simion}(2003)}]{Giuliani2003}%
  \BibitemOpen
  \bibfield  {author} {\bibinfo {author} {\bibfnamefont {G.~F.}\ \bibnamefont
  {Giuliani}}\ and\ \bibinfo {author} {\bibfnamefont {G.~E.}\ \bibnamefont
  {Simion}},\ }\href {https://doi.org/10.1016/s0038-1098(03)00573-8} {\bibfield
   {journal} {\bibinfo  {journal} {Solid State Commun.}\ }\textbf {\bibinfo
  {volume} {127}},\ \bibinfo {pages} {789} (\bibinfo {year}
  {2003})}\BibitemShut {NoStop}%
\bibitem [{\citenamefont {Simion}\ and\ \citenamefont
  {Giuliani}(2005)}]{Simion2005}%
  \BibitemOpen
  \bibfield  {author} {\bibinfo {author} {\bibfnamefont {G.~E.}\ \bibnamefont
  {Simion}}\ and\ \bibinfo {author} {\bibfnamefont {G.~F.}\ \bibnamefont
  {Giuliani}},\ }\href {https://doi.org/10.1103/PhysRevB.72.045127} {\bibfield
  {journal} {\bibinfo  {journal} {Phys. Rev. B}\ }\textbf {\bibinfo {volume}
  {72}},\ \bibinfo {pages} {045127} (\bibinfo {year} {2005})}\BibitemShut
  {NoStop}%
\bibitem [{\citenamefont {Chatterjee}\ and\ \citenamefont
  {Byczuk}(2015)}]{Chatterjee2015}%
  \BibitemOpen
  \bibfield  {author} {\bibinfo {author} {\bibfnamefont {B.}~\bibnamefont
  {Chatterjee}}\ and\ \bibinfo {author} {\bibfnamefont {K.}~\bibnamefont
  {Byczuk}},\ }\href {https://doi.org/10.1088/1742-6596/592/1/012059}
  {\bibfield  {journal} {\bibinfo  {journal} {J. Phys. Conf. Ser.}\ }\textbf
  {\bibinfo {volume} {592}},\ \bibinfo {pages} {012059} (\bibinfo {year}
  {2015})}\BibitemShut {NoStop}%
\bibitem [{\citenamefont {Chatterjee}\ \emph {et~al.}(2019)\citenamefont
  {Chatterjee}, \citenamefont {Skolimowski}, \citenamefont {Makuch},\ and\
  \citenamefont {Byczuk}}]{Chatterjee2019}%
  \BibitemOpen
  \bibfield  {author} {\bibinfo {author} {\bibfnamefont {B.}~\bibnamefont
  {Chatterjee}}, \bibinfo {author} {\bibfnamefont {J.}~\bibnamefont
  {Skolimowski}}, \bibinfo {author} {\bibfnamefont {K.}~\bibnamefont
  {Makuch}},\ and\ \bibinfo {author} {\bibfnamefont {K.}~\bibnamefont
  {Byczuk}},\ }\href {https://doi.org/10.1103/PhysRevB.100.115118} {\bibfield
  {journal} {\bibinfo  {journal} {Phys. Rev. B}\ }\textbf {\bibinfo {volume}
  {100}},\ \bibinfo {pages} {115118} (\bibinfo {year} {2019})}\BibitemShut
  {NoStop}%
\bibitem [{\citenamefont {Chatterjee}\ \emph {et~al.}(2022)\citenamefont
  {Chatterjee}, \citenamefont {Skolimowski},\ and\ \citenamefont
  {Byczuk}}]{Chatterjee2022}%
  \BibitemOpen
  \bibfield  {author} {\bibinfo {author} {\bibfnamefont {B.}~\bibnamefont
  {Chatterjee}}, \bibinfo {author} {\bibfnamefont {J.}~\bibnamefont
  {Skolimowski}},\ and\ \bibinfo {author} {\bibfnamefont {K.}~\bibnamefont
  {Byczuk}},\ }\href {https://doi.org/10.1103/PhysRevB.105.235129} {\bibfield
  {journal} {\bibinfo  {journal} {Phys. Rev. B}\ }\textbf {\bibinfo {volume}
  {105}},\ \bibinfo {pages} {235129} (\bibinfo {year} {2022})}\BibitemShut
  {NoStop}%
\bibitem [{\citenamefont {Sablikov}(2025)}]{Sablikov2025}%
  \BibitemOpen
  \bibfield  {author} {\bibinfo {author} {\bibfnamefont {V.~A.}\ \bibnamefont
  {Sablikov}},\ }\href {https://doi.org/10.1016/j.physe.2025.116213} {\bibfield
   {journal} {\bibinfo  {journal} {Physica E}\ }\textbf {\bibinfo {volume}
  {170}},\ \bibinfo {pages} {116213} (\bibinfo {year} {2025})}\BibitemShut
  {NoStop}%
\bibitem [{\citenamefont {Crommie}\ \emph {et~al.}(1993)\citenamefont
  {Crommie}, \citenamefont {Lutz},\ and\ \citenamefont {Eigler}}]{Crommie1993}%
  \BibitemOpen
  \bibfield  {author} {\bibinfo {author} {\bibfnamefont {M.~F.}\ \bibnamefont
  {Crommie}}, \bibinfo {author} {\bibfnamefont {C.~P.}\ \bibnamefont {Lutz}},\
  and\ \bibinfo {author} {\bibfnamefont {D.~M.}\ \bibnamefont {Eigler}},\
  }\href {https://doi.org/10.1038/363524a0} {\bibfield  {journal} {\bibinfo
  {journal} {Nature}\ }\textbf {\bibinfo {volume} {363}},\ \bibinfo {pages}
  {524} (\bibinfo {year} {1993})}\BibitemShut {NoStop}%
\bibitem [{\citenamefont {Hofmann}\ \emph {et~al.}(1997)\citenamefont
  {Hofmann}, \citenamefont {Briner}, \citenamefont {Doering}, \citenamefont
  {Rust}, \citenamefont {Plummer},\ and\ \citenamefont
  {Bradshaw}}]{Hofmann1997}%
  \BibitemOpen
  \bibfield  {author} {\bibinfo {author} {\bibfnamefont {P.}~\bibnamefont
  {Hofmann}}, \bibinfo {author} {\bibfnamefont {B.~G.}\ \bibnamefont {Briner}},
  \bibinfo {author} {\bibfnamefont {M.}~\bibnamefont {Doering}}, \bibinfo
  {author} {\bibfnamefont {H.-P.}\ \bibnamefont {Rust}}, \bibinfo {author}
  {\bibfnamefont {E.~W.}\ \bibnamefont {Plummer}},\ and\ \bibinfo {author}
  {\bibfnamefont {A.~M.}\ \bibnamefont {Bradshaw}},\ }\href
  {https://doi.org/10.1103/physrevlett.79.265} {\bibfield  {journal} {\bibinfo
  {journal} {Phys. Rev. Lett.}\ }\textbf {\bibinfo {volume} {79}},\ \bibinfo
  {pages} {265} (\bibinfo {year} {1997})}\BibitemShut {NoStop}%
\bibitem [{\citenamefont {Kanisawa}\ \emph {et~al.}(2001)\citenamefont
  {Kanisawa}, \citenamefont {Butcher}, \citenamefont {Yamaguchi},\ and\
  \citenamefont {Hirayama}}]{Kanisawa2001}%
  \BibitemOpen
  \bibfield  {author} {\bibinfo {author} {\bibfnamefont {K.}~\bibnamefont
  {Kanisawa}}, \bibinfo {author} {\bibfnamefont {M.~J.}\ \bibnamefont
  {Butcher}}, \bibinfo {author} {\bibfnamefont {H.}~\bibnamefont {Yamaguchi}},\
  and\ \bibinfo {author} {\bibfnamefont {Y.}~\bibnamefont {Hirayama}},\ }\href
  {https://doi.org/10.1103/physrevlett.86.3384} {\bibfield  {journal} {\bibinfo
   {journal} {Phys. Rev. Lett.}\ }\textbf {\bibinfo {volume} {86}},\ \bibinfo
  {pages} {3384} (\bibinfo {year} {2001})}\BibitemShut {NoStop}%
\bibitem [{\citenamefont {Ono}\ \emph {et~al.}(2009)\citenamefont {Ono},
  \citenamefont {Nishio}, \citenamefont {An}, \citenamefont {Eguchi},\ and\
  \citenamefont {Hasegawa}}]{Ono2009}%
  \BibitemOpen
  \bibfield  {author} {\bibinfo {author} {\bibfnamefont {M.}~\bibnamefont
  {Ono}}, \bibinfo {author} {\bibfnamefont {T.}~\bibnamefont {Nishio}},
  \bibinfo {author} {\bibfnamefont {T.}~\bibnamefont {An}}, \bibinfo {author}
  {\bibfnamefont {T.}~\bibnamefont {Eguchi}},\ and\ \bibinfo {author}
  {\bibfnamefont {Y.}~\bibnamefont {Hasegawa}},\ }\href
  {https://doi.org/10.1016/j.apsusc.2009.07.023} {\bibfield  {journal}
  {\bibinfo  {journal} {Appl. Surf. Sci.}\ }\textbf {\bibinfo {volume} {256}},\
  \bibinfo {pages} {469} (\bibinfo {year} {2009})}\BibitemShut {NoStop}%
\bibitem [{\citenamefont {Dutreix}\ \emph {et~al.}(2019)\citenamefont
  {Dutreix}, \citenamefont {González-Herrero}, \citenamefont {Brihuega},
  \citenamefont {Katsnelson}, \citenamefont {Chapelier},\ and\ \citenamefont
  {Renard}}]{Dutreix2019}%
  \BibitemOpen
  \bibfield  {author} {\bibinfo {author} {\bibfnamefont {C.}~\bibnamefont
  {Dutreix}}, \bibinfo {author} {\bibfnamefont {H.}~\bibnamefont
  {González-Herrero}}, \bibinfo {author} {\bibfnamefont {I.}~\bibnamefont
  {Brihuega}}, \bibinfo {author} {\bibfnamefont {M.~I.}\ \bibnamefont
  {Katsnelson}}, \bibinfo {author} {\bibfnamefont {C.}~\bibnamefont
  {Chapelier}},\ and\ \bibinfo {author} {\bibfnamefont {V.~T.}\ \bibnamefont
  {Renard}},\ }\href {https://doi.org/10.1038/s41586-019-1613-5} {\bibfield
  {journal} {\bibinfo  {journal} {Nature}\ }\textbf {\bibinfo {volume} {574}},\
  \bibinfo {pages} {219} (\bibinfo {year} {2019})}\BibitemShut {NoStop}%
\bibitem [{\citenamefont {Chen}\ \emph {et~al.}(2021)\citenamefont {Chen},
  \citenamefont {Duan}, \citenamefont {Fan}, \citenamefont {Hong},
  \citenamefont {Chen}, \citenamefont {Yang}, \citenamefont {Li}, \citenamefont
  {Luo},\ and\ \citenamefont {Wen}}]{Chen2021}%
  \BibitemOpen
  \bibfield  {author} {\bibinfo {author} {\bibfnamefont {X.}~\bibnamefont
  {Chen}}, \bibinfo {author} {\bibfnamefont {W.}~\bibnamefont {Duan}}, \bibinfo
  {author} {\bibfnamefont {X.}~\bibnamefont {Fan}}, \bibinfo {author}
  {\bibfnamefont {W.}~\bibnamefont {Hong}}, \bibinfo {author} {\bibfnamefont
  {K.}~\bibnamefont {Chen}}, \bibinfo {author} {\bibfnamefont {H.}~\bibnamefont
  {Yang}}, \bibinfo {author} {\bibfnamefont {S.}~\bibnamefont {Li}}, \bibinfo
  {author} {\bibfnamefont {H.}~\bibnamefont {Luo}},\ and\ \bibinfo {author}
  {\bibfnamefont {H.-H.}\ \bibnamefont {Wen}},\ }\href
  {https://doi.org/10.1103/PhysRevLett.126.257002} {\bibfield  {journal}
  {\bibinfo  {journal} {Phys. Rev. Lett.}\ }\textbf {\bibinfo {volume} {126}},\
  \bibinfo {pages} {257002} (\bibinfo {year} {2021})}\BibitemShut {NoStop}%
\bibitem [{\citenamefont {Yin}\ \emph {et~al.}(2023)\citenamefont {Yin},
  \citenamefont {Zhou}, \citenamefont {Tong}, \citenamefont {Shi},
  \citenamefont {Qin},\ and\ \citenamefont {He}}]{Yin2023}%
  \BibitemOpen
  \bibfield  {author} {\bibinfo {author} {\bibfnamefont {L.-J.}\ \bibnamefont
  {Yin}}, \bibinfo {author} {\bibfnamefont {Y.-Y.}\ \bibnamefont {Zhou}},
  \bibinfo {author} {\bibfnamefont {L.-H.}\ \bibnamefont {Tong}}, \bibinfo
  {author} {\bibfnamefont {L.-J.}\ \bibnamefont {Shi}}, \bibinfo {author}
  {\bibfnamefont {Z.}~\bibnamefont {Qin}},\ and\ \bibinfo {author}
  {\bibfnamefont {L.}~\bibnamefont {He}},\ }\href
  {https://doi.org/10.1103/PhysRevB.107.L041404} {\bibfield  {journal}
  {\bibinfo  {journal} {Phys. Rev. B}\ }\textbf {\bibinfo {volume} {107}},\
  \bibinfo {pages} {L041404} (\bibinfo {year} {2023})}\BibitemShut {NoStop}%
\bibitem [{\citenamefont {Mitsui}\ \emph {et~al.}(2020)\citenamefont {Mitsui},
  \citenamefont {Sakai}, \citenamefont {Li}, \citenamefont {Ueno},
  \citenamefont {Watanuki}, \citenamefont {Kobayashi}, \citenamefont {Masuda},
  \citenamefont {Seto},\ and\ \citenamefont {Akai}}]{Mitsui2020}%
  \BibitemOpen
  \bibfield  {author} {\bibinfo {author} {\bibfnamefont {T.}~\bibnamefont
  {Mitsui}}, \bibinfo {author} {\bibfnamefont {S.}~\bibnamefont {Sakai}},
  \bibinfo {author} {\bibfnamefont {S.}~\bibnamefont {Li}}, \bibinfo {author}
  {\bibfnamefont {T.}~\bibnamefont {Ueno}}, \bibinfo {author} {\bibfnamefont
  {T.}~\bibnamefont {Watanuki}}, \bibinfo {author} {\bibfnamefont
  {Y.}~\bibnamefont {Kobayashi}}, \bibinfo {author} {\bibfnamefont
  {R.}~\bibnamefont {Masuda}}, \bibinfo {author} {\bibfnamefont
  {M.}~\bibnamefont {Seto}},\ and\ \bibinfo {author} {\bibfnamefont
  {H.}~\bibnamefont {Akai}},\ }\href
  {https://doi.org/10.1103/PhysRevLett.125.236806} {\bibfield  {journal}
  {\bibinfo  {journal} {Phys. Rev. Lett.}\ }\textbf {\bibinfo {volume} {125}},\
  \bibinfo {pages} {236806} (\bibinfo {year} {2020})}\BibitemShut {NoStop}%
\bibitem [{\citenamefont {Mitsui}\ \emph {et~al.}(2021)\citenamefont {Mitsui},
  \citenamefont {Sakai}, \citenamefont {Li}, \citenamefont {Ueno},
  \citenamefont {Watanuki}, \citenamefont {Kobayashi}, \citenamefont {Masuda},
  \citenamefont {Seto},\ and\ \citenamefont {Akai}}]{Mitsui2021}%
  \BibitemOpen
  \bibfield  {author} {\bibinfo {author} {\bibfnamefont {T.}~\bibnamefont
  {Mitsui}}, \bibinfo {author} {\bibfnamefont {S.}~\bibnamefont {Sakai}},
  \bibinfo {author} {\bibfnamefont {S.}~\bibnamefont {Li}}, \bibinfo {author}
  {\bibfnamefont {T.}~\bibnamefont {Ueno}}, \bibinfo {author} {\bibfnamefont
  {T.}~\bibnamefont {Watanuki}}, \bibinfo {author} {\bibfnamefont
  {Y.}~\bibnamefont {Kobayashi}}, \bibinfo {author} {\bibfnamefont
  {R.}~\bibnamefont {Masuda}}, \bibinfo {author} {\bibfnamefont
  {M.}~\bibnamefont {Seto}},\ and\ \bibinfo {author} {\bibfnamefont
  {H.}~\bibnamefont {Akai}},\ }\href
  {https://doi.org/10.1007/s10751-021-01772-0} {\bibfield  {journal} {\bibinfo
  {journal} {Hyperfine Interact.}\ }\textbf {\bibinfo {volume} {242}},\
  \bibinfo {pages} {37} (\bibinfo {year} {2021})}\BibitemShut {NoStop}%
\bibitem [{\citenamefont {Berthier}(1978)}]{Berthier1978}%
  \BibitemOpen
  \bibfield  {author} {\bibinfo {author} {\bibfnamefont {C.}~\bibnamefont
  {Berthier}},\ }\href {https://doi.org/10.1088/0022-3719/11/4/024} {\bibfield
  {journal} {\bibinfo  {journal} {J. Phys. C: Solid State Phys.}\ }\textbf
  {\bibinfo {volume} {11}},\ \bibinfo {pages} {797} (\bibinfo {year}
  {1978})}\BibitemShut {NoStop}%
\bibitem [{\citenamefont {Yamani}\ \emph {et~al.}(2006)\citenamefont {Yamani},
  \citenamefont {Statt}, \citenamefont {MacFarlane}, \citenamefont {Liang},
  \citenamefont {Bonn},\ and\ \citenamefont {Hardy}}]{Yamani2006}%
  \BibitemOpen
  \bibfield  {author} {\bibinfo {author} {\bibfnamefont {Z.}~\bibnamefont
  {Yamani}}, \bibinfo {author} {\bibfnamefont {B.~W.}\ \bibnamefont {Statt}},
  \bibinfo {author} {\bibfnamefont {W.~A.}\ \bibnamefont {MacFarlane}},
  \bibinfo {author} {\bibfnamefont {R.}~\bibnamefont {Liang}}, \bibinfo
  {author} {\bibfnamefont {D.~A.}\ \bibnamefont {Bonn}},\ and\ \bibinfo
  {author} {\bibfnamefont {W.~N.}\ \bibnamefont {Hardy}},\ }\href
  {https://doi.org/10.1103/PhysRevB.73.212506} {\bibfield  {journal} {\bibinfo
  {journal} {Phys. Rev. B}\ }\textbf {\bibinfo {volume} {73}},\ \bibinfo
  {pages} {212506} (\bibinfo {year} {2006})}\BibitemShut {NoStop}%
\bibitem [{\citenamefont {Rouzi\`ere}\ \emph {et~al.}(2000)\citenamefont
  {Rouzi\`ere}, \citenamefont {Ravy}, \citenamefont {Pouget},\ and\
  \citenamefont {Brazovskii}}]{Rouziere2000}%
  \BibitemOpen
  \bibfield  {author} {\bibinfo {author} {\bibfnamefont {S.}~\bibnamefont
  {Rouzi\`ere}}, \bibinfo {author} {\bibfnamefont {S.}~\bibnamefont {Ravy}},
  \bibinfo {author} {\bibfnamefont {J.-P.}\ \bibnamefont {Pouget}},\ and\
  \bibinfo {author} {\bibfnamefont {S.}~\bibnamefont {Brazovskii}},\ }\href
  {https://doi.org/10.1103/PhysRevB.62.R16231} {\bibfield  {journal} {\bibinfo
  {journal} {Phys. Rev. B}\ }\textbf {\bibinfo {volume} {62}},\ \bibinfo
  {pages} {R16231} (\bibinfo {year} {2000})}\BibitemShut {NoStop}%
\bibitem [{\citenamefont {Czoschke}\ \emph {et~al.}(2005)\citenamefont
  {Czoschke}, \citenamefont {Hong}, \citenamefont {Basile},\ and\ \citenamefont
  {Chiang}}]{Czoschke2005}%
  \BibitemOpen
  \bibfield  {author} {\bibinfo {author} {\bibfnamefont {P.}~\bibnamefont
  {Czoschke}}, \bibinfo {author} {\bibfnamefont {H.}~\bibnamefont {Hong}},
  \bibinfo {author} {\bibfnamefont {L.}~\bibnamefont {Basile}},\ and\ \bibinfo
  {author} {\bibfnamefont {T.-C.}\ \bibnamefont {Chiang}},\ }\href
  {https://doi.org/10.1103/PhysRevB.72.035305} {\bibfield  {journal} {\bibinfo
  {journal} {Phys. Rev. B}\ }\textbf {\bibinfo {volume} {72}},\ \bibinfo
  {pages} {035305} (\bibinfo {year} {2005})}\BibitemShut {NoStop}%
\bibitem [{\citenamefont {Katsnelson}(2006)}]{Katsnelson2006}%
  \BibitemOpen
  \bibfield  {author} {\bibinfo {author} {\bibfnamefont {M.~I.}\ \bibnamefont
  {Katsnelson}},\ }\href {https://doi.org/10.1103/PhysRevB.74.201401}
  {\bibfield  {journal} {\bibinfo  {journal} {Phys. Rev. B}\ }\textbf {\bibinfo
  {volume} {74}},\ \bibinfo {pages} {201401} (\bibinfo {year}
  {2006})}\BibitemShut {NoStop}%
\bibitem [{\citenamefont {Gabovich}\ \emph {et~al.}(2019)\citenamefont
  {Gabovich}, \citenamefont {Li}, \citenamefont {Szymczak},\ and\ \citenamefont
  {Voitenko}}]{Gabovich2019}%
  \BibitemOpen
  \bibfield  {author} {\bibinfo {author} {\bibfnamefont {A.~M.}\ \bibnamefont
  {Gabovich}}, \bibinfo {author} {\bibfnamefont {M.~S.}\ \bibnamefont {Li}},
  \bibinfo {author} {\bibfnamefont {H.}~\bibnamefont {Szymczak}},\ and\
  \bibinfo {author} {\bibfnamefont {A.~I.}\ \bibnamefont {Voitenko}},\ }\href
  {https://doi.org/10.1016/j.elstat.2019.103377} {\bibfield  {journal}
  {\bibinfo  {journal} {J. Electrostat.}\ }\textbf {\bibinfo {volume} {102}},\
  \bibinfo {pages} {103377} (\bibinfo {year} {2019})}\BibitemShut {NoStop}%
\bibitem [{\citenamefont {Vangara}\ \emph {et~al.}(2019)\citenamefont
  {Vangara}, \citenamefont {Stoltzfus}, \citenamefont {York}, \citenamefont
  {Swol},\ and\ \citenamefont {Petsev}}]{Vangara2019}%
  \BibitemOpen
  \bibfield  {author} {\bibinfo {author} {\bibfnamefont {R.}~\bibnamefont
  {Vangara}}, \bibinfo {author} {\bibfnamefont {K.}~\bibnamefont {Stoltzfus}},
  \bibinfo {author} {\bibfnamefont {M.~R.}\ \bibnamefont {York}}, \bibinfo
  {author} {\bibfnamefont {F.~v.}\ \bibnamefont {Swol}},\ and\ \bibinfo
  {author} {\bibfnamefont {D.~N.}\ \bibnamefont {Petsev}},\ }\href
  {https://doi.org/10.1088/2053-1591/ab2791} {\bibfield  {journal} {\bibinfo
  {journal} {Mater. Res. Express}\ }\textbf {\bibinfo {volume} {6}},\ \bibinfo
  {pages} {086331} (\bibinfo {year} {2019})}\BibitemShut {NoStop}%
\bibitem [{\citenamefont {Zhang}\ \emph {et~al.}(2022)\citenamefont {Zhang},
  \citenamefont {Wang}, \citenamefont {Muniz}, \citenamefont {Panagiotopoulos},
  \citenamefont {Car},\ and\ \citenamefont {E}}]{Zhang2022}%
  \BibitemOpen
  \bibfield  {author} {\bibinfo {author} {\bibfnamefont {L.}~\bibnamefont
  {Zhang}}, \bibinfo {author} {\bibfnamefont {H.}~\bibnamefont {Wang}},
  \bibinfo {author} {\bibfnamefont {M.~C.}\ \bibnamefont {Muniz}}, \bibinfo
  {author} {\bibfnamefont {A.~Z.}\ \bibnamefont {Panagiotopoulos}}, \bibinfo
  {author} {\bibfnamefont {R.}~\bibnamefont {Car}},\ and\ \bibinfo {author}
  {\bibfnamefont {W.}~\bibnamefont {E}},\ }\href
  {https://doi.org/10.1063/5.0083669} {\bibfield  {journal} {\bibinfo
  {journal} {J. Chem. Phys.}\ }\textbf {\bibinfo {volume} {156}},\ \bibinfo
  {pages} {124107} (\bibinfo {year} {2022})}\BibitemShut {NoStop}%
\bibitem [{\citenamefont {Defenu}\ \emph {et~al.}(2024)\citenamefont {Defenu},
  \citenamefont {Lerose},\ and\ \citenamefont {Pappalardi}}]{Defenu2024}%
  \BibitemOpen
  \bibfield  {author} {\bibinfo {author} {\bibfnamefont {N.}~\bibnamefont
  {Defenu}}, \bibinfo {author} {\bibfnamefont {A.}~\bibnamefont {Lerose}},\
  and\ \bibinfo {author} {\bibfnamefont {S.}~\bibnamefont {Pappalardi}},\
  }\href {https://doi.org/10.1016/j.physrep.2024.04.005} {\bibfield  {journal}
  {\bibinfo  {journal} {Phys. Rep.}\ }\textbf {\bibinfo {volume} {1074}},\
  \bibinfo {pages} {1} (\bibinfo {year} {2024})}\BibitemShut {NoStop}%
\bibitem [{\citenamefont {Radzihovsky}\ and\ \citenamefont
  {Toner}(2024)}]{Radzihovsky2024}%
  \BibitemOpen
  \bibfield  {author} {\bibinfo {author} {\bibfnamefont {L.}~\bibnamefont
  {Radzihovsky}}\ and\ \bibinfo {author} {\bibfnamefont {J.}~\bibnamefont
  {Toner}},\ }\href {https://doi.org/10.1103/PhysRevE.110.014136} {\bibfield
  {journal} {\bibinfo  {journal} {Phys. Rev. E}\ }\textbf {\bibinfo {volume}
  {110}},\ \bibinfo {pages} {014136} (\bibinfo {year} {2024})}\BibitemShut
  {NoStop}%
\bibitem [{\citenamefont {Ene}\ \emph {et~al.}(2025)\citenamefont {Ene},
  \citenamefont {Lianu},\ and\ \citenamefont {Grosu}}]{Ene2025}%
  \BibitemOpen
  \bibfield  {author} {\bibinfo {author} {\bibfnamefont {V.-M.}\ \bibnamefont
  {Ene}}, \bibinfo {author} {\bibfnamefont {I.}~\bibnamefont {Lianu}},\ and\
  \bibinfo {author} {\bibfnamefont {I.}~\bibnamefont {Grosu}},\ }\href
  {https://doi.org/10.1016/j.physleta.2024.130064} {\bibfield  {journal}
  {\bibinfo  {journal} {Phys. Lett. A}\ }\textbf {\bibinfo {volume} {529}},\
  \bibinfo {pages} {130064} (\bibinfo {year} {2025})}\BibitemShut {NoStop}%
\bibitem [{\citenamefont {Hwang}\ and\ \citenamefont
  {Das~Sarma}(2007)}]{Hwang2007}%
  \BibitemOpen
  \bibfield  {author} {\bibinfo {author} {\bibfnamefont {E.~H.}\ \bibnamefont
  {Hwang}}\ and\ \bibinfo {author} {\bibfnamefont {S.}~\bibnamefont
  {Das~Sarma}},\ }\href {https://doi.org/10.1103/PhysRevB.75.205418} {\bibfield
   {journal} {\bibinfo  {journal} {Phys. Rev. B}\ }\textbf {\bibinfo {volume}
  {75}},\ \bibinfo {pages} {205418} (\bibinfo {year} {2007})}\BibitemShut
  {NoStop}%
\bibitem [{\citenamefont {Gradshteyn}\ and\ \citenamefont
  {Ryzhik}(2007)}]{Gradshteyn2007}%
  \BibitemOpen
  \bibfield  {author} {\bibinfo {author} {\bibfnamefont {I.}~\bibnamefont
  {Gradshteyn}}\ and\ \bibinfo {author} {\bibfnamefont {I.}~\bibnamefont
  {Ryzhik}},\ }\href
  {https://doi.org/https://doi.org/10.1016/B978-0-08-047111-2.50003-0} {\emph
  {\bibinfo {title} {Table of Integrals, Series, and Products}}},\ \bibinfo
  {edition} {seventh edition}\ ed.,\ edited by\ \bibinfo {editor}
  {\bibfnamefont {A.}~\bibnamefont {Jeffrey}}\ and\ \bibinfo {editor}
  {\bibfnamefont {D.}~\bibnamefont {Zwillinger}}\ (\bibinfo  {publisher}
  {Academic Press},\ \bibinfo {address} {Boston},\ \bibinfo {year}
  {2007})\BibitemShut {NoStop}%
\bibitem [{\citenamefont {Kraut}(2007)}]{Kraut2007}%
  \BibitemOpen
  \bibfield  {author} {\bibinfo {author} {\bibfnamefont {E.~A.}\ \bibnamefont
  {Kraut}},\ }\href@noop {} {\emph {\bibinfo {title} {Fundamentals of
  Mathematical Physics}}}\ (\bibinfo  {publisher} {Dover Publications},\
  \bibinfo {year} {2007})\BibitemShut {NoStop}%
\bibitem [{\citenamefont {R\"osner}\ \emph {et~al.}(2015)\citenamefont
  {R\"osner}, \citenamefont {\ifmmode \mbox{\c{S}}\else \c{S}\fi{}a\ifmmode
  \mbox{\c{s}}\else \c{s}\fi{}\ifmmode \imath \else \i
  \fi{}o\ifmmode~\breve{g}\else \u{g}\fi{}lu}, \citenamefont {Friedrich},
  \citenamefont {Bl\"ugel},\ and\ \citenamefont {Wehling}}]{Roesner2015}%
  \BibitemOpen
  \bibfield  {author} {\bibinfo {author} {\bibfnamefont {M.}~\bibnamefont
  {R\"osner}}, \bibinfo {author} {\bibfnamefont {E.}~\bibnamefont {\ifmmode
  \mbox{\c{S}}\else \c{S}\fi{}a\ifmmode \mbox{\c{s}}\else \c{s}\fi{}\ifmmode
  \imath \else \i \fi{}o\ifmmode~\breve{g}\else \u{g}\fi{}lu}}, \bibinfo
  {author} {\bibfnamefont {C.}~\bibnamefont {Friedrich}}, \bibinfo {author}
  {\bibfnamefont {S.}~\bibnamefont {Bl\"ugel}},\ and\ \bibinfo {author}
  {\bibfnamefont {T.~O.}\ \bibnamefont {Wehling}},\ }\href
  {https://doi.org/10.1103/PhysRevB.92.085102} {\bibfield  {journal} {\bibinfo
  {journal} {Phys. Rev. B}\ }\textbf {\bibinfo {volume} {92}},\ \bibinfo
  {pages} {085102} (\bibinfo {year} {2015})}\BibitemShut {NoStop}%
\bibitem [{\citenamefont {Min}\ \emph {et~al.}(2012)\citenamefont {Min},
  \citenamefont {Hwang},\ and\ \citenamefont {Das~Sarma}}]{Min2012}%
  \BibitemOpen
  \bibfield  {author} {\bibinfo {author} {\bibfnamefont {H.}~\bibnamefont
  {Min}}, \bibinfo {author} {\bibfnamefont {E.~H.}\ \bibnamefont {Hwang}},\
  and\ \bibinfo {author} {\bibfnamefont {S.}~\bibnamefont {Das~Sarma}},\ }\href
  {https://doi.org/10.1103/PhysRevB.86.081402} {\bibfield  {journal} {\bibinfo
  {journal} {Phys. Rev. B}\ }\textbf {\bibinfo {volume} {86}},\ \bibinfo
  {pages} {081402} (\bibinfo {year} {2012})}\BibitemShut {NoStop}%
\bibitem [{\citenamefont {Chen}\ and\ \citenamefont {Xu}(2015)}]{Chen2015}%
  \BibitemOpen
  \bibfield  {author} {\bibinfo {author} {\bibfnamefont {J.}~\bibnamefont
  {Chen}}\ and\ \bibinfo {author} {\bibfnamefont {H.-Z.}\ \bibnamefont {Xu}},\
  }\href {https://doi.org/10.1088/0253-6102/64/1/108} {\bibfield  {journal}
  {\bibinfo  {journal} {Commun. Theor. Phys.}\ }\textbf {\bibinfo {volume}
  {64}},\ \bibinfo {pages} {108} (\bibinfo {year} {2015})}\BibitemShut
  {NoStop}%
\bibitem [{\citenamefont {Luke}(1969)}]{Luke1969}%
  \BibitemOpen
  \bibfield  {author} {\bibinfo {author} {\bibfnamefont {Y.}~\bibnamefont
  {Luke}},\ }\href {https://books.google.ro/books?id=vO9QAAAAMAAJ} {\emph
  {\bibinfo {title} {The Special Functions and Their Approximations}}},\
  \bibinfo {series} {Mathematics in science and engineering}\ No.\ \bibinfo
  {number} {v. 1}\ (\bibinfo  {publisher} {Academic Press},\ \bibinfo {year}
  {1969})\BibitemShut {NoStop}%
\end{thebibliography}%
\end{document}